\documentclass[pre,reprint,nofootinbib,superscriptaddress]{revtex4-2}
\pdfoutput=1

\usepackage{amssymb, amsmath, amsmath, amsfonts}
\usepackage[colorlinks,linkcolor=blue,urlcolor=blue,citecolor=blue]{hyperref}
\usepackage{graphics}
\usepackage{physics}
\graphicspath{{figs/}}

\usepackage[english]{babel}

\renewcommand{\S}{\ensuremath{\mathcal{S}}\xspace}
\newcommand{\F}{\ensuremath{\mathcal{F}}\xspace}
\newcommand{\W}{\ensuremath{\mathcal{W}}\xspace}
\newcommand{\E}{\ensuremath{\mathcal{E}}\xspace}

\newcommand{\superF}{{\scriptscriptstyle(\mathcal{F})\xspace}}
\newcommand{\superW}{{\scriptscriptstyle(\mathcal{W})\xspace}}
\newcommand{\superE}{{\scriptscriptstyle(\mathcal{E})\xspace}}

\usepackage{xcolor}
\usepackage{xspace}
\usepackage{ifthen}

\newcommand{\carlo}[1]
{\ifthenelse{\equal{\showcomments}{true}}
{{\color{blue}{\textbf{Carlo says:} #1}}}{\xspace}}

\newcommand{\andrea}[1]
{\ifthenelse{\equal{\showcomments}{true}}
{{\color{orange}{\textbf{Andrea says:} #1}}}{\xspace}}

\newcommand{\showcomments}{true}

\begin{document}

\title{Relational Quantum Mechanics is about Facts, not States:\\[1mm]{A reply to Pienaar and Brukner}}

\author{Andrea {Di Biagio}}
\email{andrea.dibiagio@uniroma1.it}
\affiliation{Dipartimento di Fisica, Sapienza Universit\`a di Roma, I-00185 Roma, Italy}
\author{Carlo Rovelli}
\affiliation{Aix-Marseille University, Universit\'e de Toulon, CPT-CNRS, F-13288 Marseille, France}
\affiliation{Department of Philosophy and the Rotman Institute of Philosophy, 1151 Richmond St.~N London  N6A5B7, Canada}
\affiliation{Perimeter Institute, 31 Caroline Street N, Waterloo ON, N2L2Y5, Canada}

 \begin{abstract}
\noindent In recent works, \v{C}aslav Brukner and Jacques Pienaar have raised interesting objections to the relational interpretation of quantum mechanics. We answer these objections in detail and show that, far from questioning the viability of the interpretation, they sharpen and clarify it. 
\end{abstract}

\maketitle

\section{Introduction}
\label{intro}

Two recent papers, one by Jacques Pienaar \cite{pienaar2021quintet} and one by \v{C}aslav Brukner \cite{brukner2021qubits}, present insightful observations and objections on the Relational interpretation of Quantum Mechanics (RQM, also known as relational quantum mechanics) \cite{rovelli1996relational, rovelli2018space, laudisa2019relational, dibiagio2021stable}.   Here we discuss these papers in detail.  We point out that the observations in them are not challenges against RQM: they are arguments that clarify and sharpen some aspects of this interpretation.  Since Pienaar's paper is more detailed, we mainly address it, mentioning Brukner's paper where relevant. 

Pienaar separates his objections to the relational interpretation into two parts.  The first regards the analogy between RQM and special relativity; the second regards the status of objectivity in RQM. In the first part, Pienaar points out that the analogy with special relativity is only partial: the sense in which variables are ``relative" in special relativity is more restricted than the sense in which variables are ``relative" in RQM.  In the second part, he argues that RQM cannot be reduced to the relativity of variables, because facts themselves are relative, and there is no absolute way of comparing the perspectives of two systems.    

Both observations are correct. But they are not objections to RQM. They are considerations that emphasise the radicality of the RQM perspective.  RQM does not pretend to make quantum theory less revolutionary than what it is.  It only claims that there exists a coherent and complete way of thinking about quantum phenomena that makes sense without requiring many worlds, hidden variables, cognitive agents, or a macroscopic classical world.  Hence the two objections by Pienaar are only objections to the hope to spoil RQM of its core (radical) idea. 

Pienaar makes his objections concrete in the form of five ``no-go theorems" which are supposed to pitch the claims of RQM against one another.  To do so, he summarises RQM in terms of six ``claims", that he names \textbf{RQM:1}--\textbf{RQM:6}.  This is a detailed and mostly accurate account of RQM. But it contains one misstep: a misrepresentation of the claims \textbf{RQM:5} and \textbf{RQM:6} (see below).  This misrepresentation is common to both \cite{brukner2021qubits} and \cite{pienaar2021quintet}, and regards the meaning of the quantum state.  In RQM, the quantum state is not a representation of reality: it is always a relative state and is only a mathematical tool used to predict probabilities of events \emph{relative to a given system}. The quantum state of a composite system relative to an external system is not an account or record of relative events \emph{between the subsystems} of the composite system. It is only a mathematical tool useful for predicting probabilities of events \emph{relative to the external system}.  Assuming that the quantum state is more than this is the misunderstanding leading to the apparent contradictions.  

This same mischaracterisation of RQM undermines Brukner's critique in  \cite{brukner2021qubits}.
Brukner's theorem then does not appear as a critique of RQM. It becomes instead a restriction on the concept of knowledge---concept that plays no \textit{fundamental} role in the formulation of RQM.

Because of the mischaracterisation of \textbf{RQM:5} and \textbf{RQM:6}, and the consequent over-emphasis on the quantum state, the theorems, as we shall see, either fall apart or become evidence of the \emph{consistency} of the interpretation. Pienaar's and Brukner's acute arguments actually turn out to illuminate and emphasise the consistency of the interpretation, rather than challenging it.

In section \ref{sec:axiomstheorems}, we comment on Pienaar's formulation RQM's  claims, pointing out where it is imprecise. We also briefly anticipate how each of the five no-go theorems is resolved in a proper understanding of RQM.  In section \ref{sec:relativity}, we comment on the relativistic analogy and, in section \ref{sec:objectivity}, we address Pienaar's comments about objectivity in RQM. In this context, we present also a general philosophical consideration regarding the physical meaning of a subject's knowledge. In section \ref{sec:qubits}, we respond to Brukner's paper. 

As in Pienaar's article, we consider three interacting systems \W, \F and \S, and describe the events relative to either \F or \W. The notation is meant to suggest the setup of Wigner's friend thought experiment \cite{wigner1962remarks}.

\vfill

\section{Claims and theorems}
\label{sec:axiomstheorems}

Pienaar summarizes the RQM literature in terms of six claims, reported in full for reference:
\begin{description}
\addtolength{\itemsep}{-1mm}

\item[\rm\textbf{RQM}]\textbf{\!\!1 Any system can be an observer.}  Any physical system can play the role of an observer in a physical interaction.

\item[\rm\textbf{RQM}]\textbf{\!\!2 No hidden variables.} Any variable that exists in the observer's causal past and which is relevant to predictions about future quantum events relative to the observer must be a quantum event contained in their perspective.

\item[\rm\textbf{RQM}]\textbf{\!\!3 Relations are intrinsic.} The relation between any two systems $\mathcal{A}$ and $\mathcal{B}$ is independent of anything that happens outside these systems' perspectives. In particular, the state of $\mathcal{B}$ relative to $\mathcal{A}$ depends only upon $\mathcal{A}$'s observation of $\mathcal{B}$ and $\mathcal{A}$'s past history of interactions (similarly for the state of $\mathcal{A}$ relative to $\mathcal{B}$). 

\item[\rm\textbf{RQM}]\textbf{\!\!4 Comparisons are relative to one observer.} It is meaningless to compare the accounts of any two observers except by invoking a third observer relative to which the comparison is made.

\item[\rm\textbf{RQM}]\textbf{\!\!5 Any physical correlation is a measurement.} Suppose an observer measures a pair of systems and thereby assigns them a joint state which exhibits perfect correlations between some physical variables. Then the two systems have measured each other (entered into a measurement interaction) relative to the observer, and the physical variables play the roles of the `pointer variable' and `measured variable' of the systems.

\item[\rm\textbf{RQM}]\textbf{\!\!6 Shared facts.} In the Wigner's friend scenario, if $\mathcal{W}$ measures $\mathcal{F}$ to `check the reading' of a pointer variable (i.e.\ by measuring $\mathcal{F}$ in the appropriate `pointer basis'), the value he finds is necessarily equal to the value that $\mathcal{F}$ recorded in her account of her earlier measurement of $\mathcal{S}$.

\end{description}

This is a good summary of RQM, but some points are slightly misleading, and one is strongly misleading.  Let us comment on each claim.

\textbf{RQM:1 Any system can be an observer} is essentially correct but poorly phrased, because of the term ``observer".   RQM distinguishes relative facts from stable facts  \cite{dibiagio2021stable}. Relative facts (or ``events") form the basis of the  ontology; they are ubiquitous and do not require any special property of the physical systems involved in order to happen.  Stable facts are facts stabilised by decoherence, in the sense that their relativity can be ignored by a large class of systems \cite{dibiagio2021stable}.  It is better to reserve the use of operational expressions such as ``observer" and ``measurement" to those specific situations where there is enough decoherence to underpin stability, for instance, when there is a scientist making observations, or a macroscopic system storing memory.%
\footnote{To be sure, the RQM literature does use the operational terminology ambiguously and it is indeed common to call ``observer" any system with respect to which a variable takes values.  We shall also indulge in this abuse of language below.}
Terminology aside, the actual content of \textbf{RQM:1} is that we assume something can happen relative to any system---not only measuring apparata or  ``observers" that are special in any sense. 

So we would rephrase this claim as:
\begin{description}
\addtolength{\itemsep}{-2mm}
\item[\rm\textbf{RQM}]\textbf{\!\!1$\star$ Events, or facts, can happen relative to any physical system.} Events happen in interactions between any two systems and can be described as the actualisation of the value of a variable of one system relative to the other.  
\end{description}

\textbf{RQM:2 No hidden variables} is a statement about the universality of QM.
It is correct, but RQM is consistent with the time-reversal invariance of fundamental physics (see \cite{dibiagio2021arrow}), and thus the formulation given by Pienaar must be generalised: it remains valid when swapping `past' and `future'. \textbf{RQM:3 Relations are intrinsic} also does not require any modification.

\textbf{RQM:4 Comparisons are relative to one observer} is another key tenet of RQM.  The idea is that contradictions arise when trying to equate descriptions of physics in two different contexts, namely relative to different systems.  This is for instance what happens in the  Frauchiger and Renner experiment \cite{frauchiger2018quantum}. See \cite{dibiagio2021stable} for an analysis of this situation.  We rephrase this claim in a cleaner language as: 

\begin{description}
\addtolength{\itemsep}{-2mm}
\item[\rm\textbf{RQM}]\textbf{\!\!4$\star$ Comparisons are only relative to a system.} It is meaningless to compare events relative to different systems, unless this is done relative to 
a (possibly third) system.
\end{description}
The point is that comparisons can only be made by a (quantum-mechanical) interaction. In the Wigner's friend setup, \W might compare the result of his measurement on \S with that of \F only by \emph{physically} interacting with \F in an appropriate manner. There is no meaning in comparing facts relative to \W's with facts relative to \F's, (or relative to Schr\"{o}dinger and his cat) apart from this direct physical interaction.

We now come to the troublesome points.  \textbf{RQM:5 Any physical correlation is a measurement} is the main problem with Pienaar's account.  In RQM, facts determine states, not the other way around.  Knowing the state of a system \S is not sufficient to deduce the set of facts relative to the subsystems of \S.  This is in fact precisely what Pienaar's theorems 3 and 5 show.

The problem here is what determines what.  Pienaar and Brukner take the state as primitive and assume that out of the state one can deduce which events happen in a composite system. This is not RQM. In RQM, it is the other way around. Events are primitive. Their happening is partially reflected in the state of the composite system relative to a third system. But only partially. Events cannot be read out of the state.  

The existence of a correlation between two variables gives indications about events, but in general it is not sufficient to tell which event was or was not realised.  To know what event lead to the creation of a correlation, one needs to know more, for example the dynamics that coupled the two systems and, in particular, what variables are involved in the interaction.

Besides this key misrepresentation, there is also a terminological problem in \textbf{RQM:5}, parallel to the one pointed out for \textbf{RQM:1}.  Pienaar calls a ``measurement'' what the RQM literature calls an event that establishes a fact. It is much better to reserve the loaded expression ``measurement'' to interactions that stabilise certain facts and require decoherence. 

A proper reformulation of \textbf{RQM:5}, is:

\begin{description}
\addtolength{\itemsep}{-2mm}
\item[\rm\textbf{RQM}]\textbf{\!\!5$\star$  An interaction between two systems results in a correlation within the interactions between these two systems and a third one.}   With respect to a third system \W, the interaction between the two systems \S and \F is described by a unitary evolution that potentially entangles the quantum states of \S and \F. 
\end{description}

As we shall see, while {\rm\textbf{RQM:5}} is in tension with  \textbf{RQM:3}, {\rm\textbf{RQM:5$\star$}} is not.

\textbf{RQM:5$\star$} goes hand in hand with \textbf{RQM:1$\star$}. These two assumptions together provide the resolution of the measurement problem in RQM.   Von Neumann measurements are compatible with unitary evolution because they describes facts relative to two interacting systems (\S and \F) while the unitary evolution regards facts relative to a third system (\W).

Finally, \textbf{RQM:6 Shared facts} as stated by Pienaar is either wrong (if it is intended to override \textbf{RQM:4}) or a tautology. It is not possible to decide which because Pienaar does not mention the context of the comparison. According to \textbf{RQM:4}, the \emph{only} meaning of a comparison between an event relative to \F and an event relative to \W is in the context of a measurement made by a specified system.

A non ambiguous claim is:
\begin{quote}
    \textbf{RQM:6$\star$ Shared facts.} In the Wigner's friend scenario, if $\mathcal{W}$ measures \S on the same basis on which \F did, then appropriately interacts with $\mathcal{F}$ to `check the reading' of a pointer variable (i.e.\ by measuring $\mathcal{F}$ in the appropriate `pointer basis'), the two values found are in agreement.
\end{quote}

We briefly anticipate the resolution of the ``no-go" theorems, discussed in detail in the following sections. 
\begin{itemize}
    \item Theorem 1 does not bite because it relies on Pienaar's version of \textbf{RQM:5}.
    \item  Theorem 2 relies on two assumptions that are not valid in RQM because they misrepresent the role of the quantum state in the interpretation.
    \item  Theorem 3 relies on \textbf{RQM:6} which is incorrect.
    \item Theorem 4  does not bite because of \textbf{RQM:5} again.
    \item Theorem 5 relies on \textbf{RQM:5}, which is incorrect.
\end{itemize}
Theorems 2, 3, and 5 offer two alternatives (two `horns'). As we shall discuss below, RQM `grabs a horn' in each of them. Theorem 2 elucidates what RQM is about, while grabbing the horn in theorems 3 and 5 simply amounts to correcting Pienaar's mischaracterisation of RQM. Theorems 1 and 4 do not apply to RQM, for the same important reason, they are based on the misunderstanding of the role of the quantum state.

\section{The analogy with relativity}
\label{sec:relativity}

The analogy between SR and RQM is often used in presentations of RQM.  Pienaar shows in detail that the relationalism on which RQM is based is far more radical that the relationalism that underpins classical relativity. Therefore the conceptual novelty of quantum theory cannot be reduced to a simple recognition that all variables are relative, like velocity is relative in mechanics.  

Pienaar's judgement is spot on.  He characterises the relationality of RQM with the slogan ``facts are relative".  He is right in this.  He suggests changing the name of the interpretation to `Relative-facts interpretation of Quantum Mechanics'. That might be appropriate, but `Relational' also works, because reality relative to one system---the collections of facts relative to that system---is composed of direct interactions this system has with the rest of the world.  Rendering facts relative is a generalisation of relativity, albeit a drastic one.  

On the other hand, Pienaar's claim that ``without the conceptual analogy to classical relativistic relations, RQM would lose its core motivation as an interpretation" is a non sequitur.  The interest and the value of RQM does not  depend  on it being analogous to something else.  As any interpretation of quantum mechanics, it derives its worth from the extent it elucidates our quantum world. 

In addition, there are two other aspects of the analogy, that Pienaar disregards.  First, special relativity is a conceptual advance based on the realisation that a previously ``obvious'' notion---absolute simultaneity---is in fact inappropriate to describe the world. RQM is a conceptual advance based on the realisation that another previously ``obvious'' notion---absolute facts---may in fact be inappropriate to describe the world. (We might soon have empirical evidence for this, see~\cite{bong2020strong}.)

Second, there is a \textit{methodological} similarity between RQM and special relativity: the idea of searching for transparent physical principles from which the mathematical structure of the theory can be defined. The two principles proposed in the first paper on RQM \cite{rovelli1996relational}, are
\begin{enumerate}
    \item The relevant information that can be extracted from a finite region of the phase space of a physical system is finite,
    \item It is always possible to extract novel relevant information from a physical system.
\end{enumerate}
are based on the idea that the theory describes the relative information that a system can gather about another system. These principles serve as the first two axioms of H\"ohn's and Wever's compelling reconstruction \cite{hohn2017toolbox,hohn2017quantumb,hohn2017quantum}. 

In brief, the analogy with relativity played a historical role in the development of RQM and has some interest despite the fact that it is not complete. 
Perhaps Pienaar hoped that the radical conceptual novelty of quantum mechanics could be reduced to nothing else than some minor extension of special relativity.   If the literature on RQM has given this impression, this is a mistake.  RQM is genuinely radical.

Let us now look at the two theorems with which Pienaar supports his claim. 

\subsubsection*{No-go theorem 1}
\label{sec:nogo1}

\begin{quote}
    \textbf{Dilemma:} Suppose a system $\mathcal{F}$ has measured $\mathcal{S}$, and this fact is verified by a third system $\mathcal{W}$ who measures $\mathcal{F}$-$\mathcal{S}$. Then there exist situations in which one of the following must be true:\\
(i) $\mathcal{F}$ has measured $\mathcal{S}$ simultaneously in incompatible bases, relative to $\mathcal{W}$;\\
(ii) The basis in which $\mathcal{F}$ has measured $\mathcal{S}$ is indeterminate relative to $\mathcal{W}$.
\end{quote}

Pienaar understands this dilemma to be a no-go theorem because both alternatives contradict some of the RQM claims.   In particular, (ii) contradicts {\bf RQM:5}.  The solution of the difficulty is that (ii) is correct and does not contradict any of the RQM claims, because it is {\bf RQM:5$\star$} and not {\bf RQM:5} that characterises RQM and (ii) is not in contradiction with  {\bf RQM:5$\star$}. Underlying this, there is a misunderstanding of the role of the quantum state in RQM.

Let us see this in more detail.
In the proof of the dilemma, a situation is considered in which the state of $\S-\F$ relative to \W is
\begin{equation}\label{state-x}
  \ket{\Psi}_{\S\F} = \sum_i   \alpha_{i}   \ket{x_i}_{\S} \ket{Fx_{i}}_{\F},
\end{equation}
where $\{x_i\}$ and $\{Fx_i\}$ denote eigenvalues of some observables $X$ and $F_X$ of \S and \F respectively. 
Pienaar notes that in general this Schmidt decomposition is not unique, and one could find other observables $Y$ and $F_Y$ such that
\begin{equation}\label{state-y}
  \ket{\Psi}_{\S\F} = \sum_n  \beta_{n}\ket{y_n}_{\S}  \ket{Fy_{n}}_{\F} .
\end{equation}
He then uses of \textbf{RQM:5} (every correlation is a measurement) to derive horn (i) of the dilemma. Since there is a correlation both between $X$ and $F_X$ and between $Y$ and $F_Y$, then, allegedly, \F has measured \S simultaneously on the incompatible bases $X$ and $Y$. This is not the case in RQM, \textbf{RQM:5} cannot be applied. All that $\ket{\Psi}_{\S\F}$ tells us is which kinds of correlations exist between the variables of the two systems, relative to $W$. 

The confusion arises also because of Pienaar's use of the word `measurement'. Relative to \W, the only meaning that can be ascribed to the question of whether or not \F has ``measured" \S is whether there is a correlation between the relevant variables of the two systems.  Since there is a correlation between different pairs of variables, \emph{in this sense and only in this sense}, the ``measurement" happened in multiple bases.  The strangeness of the statement is only the inappropriate use of the expression ``measurement" in this situation.  As Pienaar himself notes, the RQM literature often recommends not to use this misleading expression. If we use proper expressions, everything returns to reasonable.   With respect to \W, is there a correlation between variables of \S and variables of \F? Yes there is. In which basis? In more than one basis. 

So how do we know which of \S's variables became definite relative to \F? We do not, if we only know the state $\ket{\Psi}_{\S\F}$. More information can be obtained from the dynamics of the system.  The state \eqref{state-x} for instance may arise as a result of an interaction between \S and \F in which the evolution of \F depends on the value of the variable $X$ of \S. For example, the interaction Hamiltonian can depends on this variable.  From the perspective of \F, this interaction leads to the actualization of the variable $X$ of \S. But the same state relative to \W could arise via an interaction Hamiltonian that depends on the variable $Y$, and then it is this variable that actualizes relative to \F.  The physics of the two processes is different, but results in the same state relative to \W, namely in the same probability distribution of events relative to \W.   The final state relative to \W lacks information about what happens among subsystems. 

Pienaar refers to the observable $M$ of the combined system $\S-\F$ that was introduced in \cite{rovelli1998incerto}. This is an observable that \W can measure to check the existence of a perfect correlation between certain variables:
\begin{equation}
  M\ket{x_i}_\S\ket{Fx_j}_\F = \delta_{ij}\ket{x_i}_\S\ket{Fx_i}_\F.
\end{equation}
The same $M$ can be expressed as
\begin{equation}
  M\ket{y_n}_\S\ket{Fy_m}_\F = \delta_{nm}\ket{x_n}_\S\ket{Fx_m}_\F.
\end{equation}
Measuring $M=1$ tells us that the correlation exists and is maximal. This is compatible with  \textit{either} $X$ \textit{or} $Y$ having taken a definite value relative to \F. The value of $M$ on its own, does not allow \W to know which variable is definite relative to \F. 

The central idea of RQM is that, since the only way for \W and \F to communicate is via a quantum mechanical measurement, there is no meaning to any other form of relations between the two.  Here Pienaar is  equating two distinct statements: (i) a variable of \S has a value with respect to \F, and (ii) with respect to \W, there is a correlation to be expected between a variable of \S and a pointer variable of \F.  The first implies the second, but the second does not imply the first. The second can regard multiple bases even while the first cannot.

\subsubsection*{No-go theorem 2}
\label{sec:nogo2}

This theorem is meant to express a contradiction between a set of three assumptions (i)-(iii) constraining the set of possible states that two systems \F and \W might assign to a third system \S and the fact (iv) that not all state assignments are good states assignments. We report here the two relevant assumptions: 
\begin{quote}
(ii) Any valid state assignment $\ket{\psi}_{\mathcal{S}}$ by $\mathcal{F}$ can always be verified by $\mathcal{W}$. That is, there must exists a `pointer basis' of $\mathcal{F}$ such that, if $\mathcal{W}$ were to measure in this basis and condition on the outcome, there would be a nonzero probability of updating the state of $\mathcal{S}$ relative to $\mathcal{W}$ to $\ket{\psi}_{\mathcal{S}}$.\\
(iii) Conversely, any assignment $\ket{\psi}_{\mathcal{S}}$ by $\mathcal{F}$ which can be verified by $\mathcal{W}$ (in the above sense) must be a valid possible assignment for $\mathcal{F}$.
\end{quote}
The theorem is again expressed as a dilemma:
\begin{quote}
    \textbf{Dilemma:} The set of assumptions (i)-(iii) are together incompatible with (iv). Specifically, given that $\mathcal{W}$ assigns an entangled state [$\ket{\Psi}_{\S\F}$] of the form [\eqref{state-x}], and assuming the coefficients $\alpha_i$ are all nonzero, then \textit{every} pure state in the Hilbert space of $\mathcal{S}$ is a possible state relative to $\mathcal{F}$.
\end{quote}

RQM resolves this no-go theorem by rejecting assumptions (ii) and (iii). 

Again, the point is the role of the quantum state in RQM. The state does not represent a description of reality; it is a computational tool to compute the likelihood of events. Say  \W assigns state \eqref{state-x} to $\S-\F$ and then measures $F_X$ and finds the value $Fx$. Then \W will have to update the quantum state of $\S-\F$ to $\ket{x}_\S\ket{Fx}_\F$. But in no way is \W allowed to conclude that \F had assigned the state $\ket{x}_\S$ to \S.
For \W to conclude that the new state of \S relative to them is the state that \S had assigned, \W would need to know that the variable $X$ had become a fact relative to \F.

Let us be even more explicit, and consider the original Wigner's friend thought experiment, where \F is an actual human in a lab and the operational talk of the previous paragraphs can be understood literally. Wigner knows that Friend measures a qubit on the computational basis, and that the value of $Z$ is then a fact relative to Friend. Wigner assigns a state proportional to
\begin{equation}
 \ket{0}_\S\ket{F0}_\F +   \ket{1}_\S\ket{F1}_\F
\end{equation}
to the combined system. If Wigner then measures the Friend on the $FZ$ basis and obtains $F0$, he is allowed to conclude that $F$ had assigned the state $\ket{0}$ to \S. What happens instead if Wigner decides instead to measure Friend on the complementary basis
$
  \{\ket{F\pm} \propto \ket{F0} \pm \ket{F1}\}
$
and obtains $F+$? Despite his experimental genius, he would be a fool to entertain that Friend had assigned the state $\ket+$ to \S! Wigner's choice of measuring on this complementary basis meant he had to forsake the ability to reveal Friend's assignment.

\subsubsection*{How radical is radical?}

One point in Pienaar's rhetoric is to emphasize the radical relationalism of quantum  phenomenology contrasting it with the consistency of the classical world. For instance, Pienaar writes:
\begin{quote}
    when two observers are in a situation where they disagree about the state of a system in RQM, the state relative to one observer \textit{places no non-trivial constraints on the state relative to the other observer}, in stark contradistinction to disagreements about velocity and other classical quantities in relativity.
\end{quote}
The misleading aspect of this rhetoric is that it ignores the physical source of the classical consistency. Classical consistency is not incompatible with quantum physics. On the contrary, its origin is clarified: it is the result of constant interactions and decoherence.  Because of decoherence, the world experienced by humans is extremely stable and because of the frequent interactions, stable facts ascertained by different observers are in agreement. 
Hence, in practice, facts relative to one observer \textit{do} place strict constraints on stable facts relative to another. 
This is why human creatures agree on the quantum state to assign to a system, on non-relational properties they assign to systems, and on the existence of a shared reality.

RQM does not bring any subversion to the stability and coherence of this classical, macroscopic world.  Instead, it shows that, by recognizing the ultimately relative nature of events,  we can have a coherent understanding of nature \emph{beyond} the macroscopic regime in which the approximation that facts are \emph{perfectly} stable is assumed to hold.   

Another rhetorical move by Pienaar is to compare the RQM terminology with analogous terminology in different contexts.  For instance, Pienaar writes 
\begin{quote}
Far from having de-mystified quantum mechanics by appealing to relations, RQM has merely mystified the concept of a `relation'.     
\end{quote}
RQM takes the notions of physical system and quantum events happening between systems as primary. Quantum events involve two systems, are discrete, and are described by one variable of one system taking a value relative to the other system.  The world is not described by the individual properties of individual systems, but by relative properties. These are called `relations' because they involve more than one system. There is nothing mystifying in this terminology.   `Relations' have to be intended within this  conceptual scheme,  not in the conceptual scheme of classical mechanics, where they are subsidiaries of properties of individual systems.  If Pienaar hoped that quantum theory could be understood by retaining the classical conceptual scheme and just focusing on relations within that scheme, he is comprehensibly disappointed, and comprehensibly finds the RQM notion of `relation' mystified.

All things considered, the main objection that Pienaar raises to RQM is not that it is inconsistent: it is that of being more radical than what he  hoped for. 

\section{On objectivity}
\label{sec:objectivity}

This part of Pienaar's objections have to do with the consequences of \textbf{RQM:4} have on notions of objectivity and the extent to which different perspective can be shown to agree.

\subsubsection*{No-go theorem 3}
\label{sec:nogo3}

\begin{quote}
    \textbf{Dilemma:} RQM cannot consistently maintain both the principle of \textbf{RQM:6: shared facts}, and the principle of \textbf{RQM:4: comparisons are relative to one observer}. Rejecting one or the other leads to the following two horns:\\
(i) If RQM rejects \textbf{RQM:6}, then it either implies \textit{solipsism}, or else an \textit{ontology of island universes} (these terms will be defined at the end of this section).\\
(ii) If RQM instead rejects \textbf{RQM:4}, it becomes vulnerable to our next no-go theorem, in Sec. IV B.
\end{quote}

As anticipated in section \ref{sec:axiomstheorems}, Pienaar's formulation of \textbf{RQM:6}, is loose enough that it is either wrong, or a tautology. The proof of this no-go theorem illustrates this point. 

Pienaar tries to derive the contradiction in the following way. Consider our two systems \F and \W interacting with \S. The quantum state of \S relative to \W or \F will depend on the interactions between these three systems. He proceeds:

\begin{quote}
Now suppose we have before us a description of $\mathcal{W}$'s account, and a description of $\mathcal{F}$'s account -- laid out `side by side' in a view from nowhere, so to speak -- and we would like to know: \textit{are these accounts mutually consistent?} 
\end{quote}
Pienaar correctly points out that
\begin{quote}
according to \textbf{RQM:4}, this is not a well-posed question, because there is no `view from nowhere'
\end{quote}
and yet he also holds that
\begin{quote}
\textbf{RQM:6} requires that this question be well-posed, for otherwise there would be no way to assert that two observer's accounts are in agreement.
\end{quote}
If Pienaar intended \textbf{RQM:6} to imply that there is a `view from nowhere,' from which to compare all accounts of reality, then clearly one must reject \textbf{RQM:6}, as it contradicts \textbf{RQM:4}.

Crucially, however, there \textit{is} a way to ``assert that two observer's accounts are in agreement'' (despite having rejected \textbf{RQM:6}): have \F write down her account and let \W read it and compare it with its own.\footnote{Or, in less anthropomorphised terms: let the dynamics be such that \F encodes its account of \S in a suitable pointer variable and let \W interact with that variable.} That \W will find that \F's account is in agreement with his, is precisely the content of \textbf{RQM:6$\star$}, which is clearly compatible with \textbf{RQM:4}.

Rejecting Pienaar's \textbf{RQM:6} and replacing it with \textbf{RQM:6$\star$}  amounts to grabbing horn (i) of the dilemma.
Pienaar claims that this would plunge us  into solipsism or into an ontology of island universes. Would it? 

\subsubsection*{Solipsism?}

The claim that RQM leads to ``solipsism" has appeared elsewhere, especially in popular science (see for instance \cite{gefter2014trespassing}).

In the philosophical literature and in common parlance, solipsism has nothing to with incomplete  of communication between physical systems. It is instead the idea that there is a single subject that exhausts all of reality and that the rest of reality only exists as the experience of that single subject. 

 This is exactly the opposite of RQM.  The main assumption of RQM, its defining assumption, in fact, is the antithesis of solipsism: the world is not what is perceived by a single special entity---it is a network of interactions between equal status entities.  

Pienaar does eventually concede\footnote{In fact, it is puzzling that he chooses to levy such a charge in the first place. He recently wrote an excellent comparison \cite{pienaar2021qbism} between RQM and QBism, another interpretation often accused of being solipsistic. Neither interpretation is solipsist,  for the same reason.} that probably RQM does not propose solipsism. He correctly characterises RQM's view: there are facts relative to every system, but that the different perspectives on reality, namely, the ensemble of facts relative to a single system, cannot be compared in an absolute manner; they can only be compared via a physical interaction. This is correct. 

He calls this an ontology of ``island universes.'' Fine, if he likes this name. We do not, let us explain why.

\subsubsection*{Embodied knowledge}

There is a subtle but important philosophical issue involved here. Consider on the case in which the systems \F and \W are actually ``observers" in the rich sense of the term. Say they are humans with laboratories, notebooks and books that store and process knowledge about the world.  Let us focus on \F.  What is the meaning of the statement that \F has knowledge about the world, for instance about \S? 

There are two possible answers. The first is a naturalistic answer. The second is a dualistic or idealistic answer.   
According to the first, this is a statement about the actual physical configuration of the ink and the notebooks, the charges in the computers and the synapses in the brain in \F and about the correlation of these with whatever can be observed in \S.  
According to the second, \F's knowledge is something over and above its physical configuration. In this case, the ``inaccessibility'' of \F's knowledge, namely of the ``universe as seen by \F'' is indeed there.   But this only follows because one assumes that knowledge is unphysical.  

We adhere to a naturalistic philosophy. In a naturalistic philosophy, what \F ``knows" regards  physical variables in \F.  And \emph{this} is accessible to \W.   If knowledge is physical, it is accessible by other systems via physical interactions.  
It is precisely for this reason that knowledge is also subjected to the constraints and the physical accidents due to quantum theory. A physical interaction can and does destroy knowledge, because of standard Heisenberg uncertainty. 
Hence, ultimately, the intuition that disturbs Pienaar is a residual of anti-naturalism: the idea that knowledge can remain immune from quantum phenomena, because it can be disembodied. 

\subsubsection*{Are relative facts needed?}

Clarified this (subtle) point, there remains\footnote{Brukner reported a similar concern  in an email \cite{brukner2021private}.} in Pienaar's \cite{pienaar2021quintet} an objection:
\begin{quote}
this proliferation of disjoint universes is not motivated by observations, nor does it serve any explanatory purpose.
\end{quote}
Every interpretation of quantum theory is ``motivated by observations" in the sense that it is an attempt to devise a conceptual scheme that makes sense of a vast number of observations. More precisely, to make sense of the fact that observations are well described by quantum theory. As such, it is deeply rooted in observations: without observations, quantum theory---and its tentative interpretations---would never have appeared. 

More to the point, what is the explanatory purpose of the multiplication of perspectives in RQM? 
The answer is that it offers a possible explanation to the key mystery of quantum physics: the apparent special role that ``observers" seem to have in the theory. 
 
 RQM illuminates this mystery by denying that there is anything special in observers, in the following general sense: facts happen relative to any system (\textbf{RQM:1}).   What \textit{is} special in a (large class) of macroscopic observers is only that decoherence and frequent interactions stabilise and render consistent for them many relative facts.  RQM is the observation that quantum physics can be made sense of also beyond the limit of perfect decoherence. 
  
Thus, the ``explanatory purpose'' of RQM's multiplication of perspectives (the idea that facts happen at interactions between any two systems) is that it serves as a possible solution to the measurement problem. It helps to answer questions like: 
\begin{itemize}
\item Q: When does something become a fact?
\\ A: Something becomes a fact, relative to you, when you interact with a system. 
\item Q: How does Schr\"{o}dinger's cat feel? \\ A: Either awake looking at the vial, or asleep having a dream. The cat does not stop having experiences only because the box is sealed off from the rest of the lab.
\item  Q: What physical systems are measuring apparata? \\ A: Any system whose pointer variables (i) get appropriately entangled to a variable you are interested and (ii) with which you can interact. 
\item Q: When does the wavefunction collapse? \\ A: The wavefunction for \S relative to \W collapses whenever \W interacts with---gets a kick from---\S and therefore \W gathers information about \S.
\end{itemize}

\subsubsection*{Island universes}

The expression ``island universes'' that Pienaar uses to RQM's discredit is taken  from Huxley's \textit{The Doors of Perception} \cite{huxley2004doors}, where ``island universes'' is applied to  conscious experiences. The situation with conscious experiences is in fact analogous to that in quantum physics, but instead of weakening the motivation for postulating multiple perspectives, it strengthens it.

Do I have direct evidence that other humans have a first-person experience of reality like mine?  I do not. Is thinking that other humans have experiences like mine a hypothesis that is ``not motivated by observations, nor [serves] any explanatory purpose''?  Of course not!   The alternative is to think that I myself am the only conscious being in the universe, namely solipsism!   

We have ample reasons to believe that we share conscious experiences with (at least) other humans. By the same token, RQM points out that we have reasons to believe that we share the reality of  perspectival facts with any physical system.

This is the core of RQM: I understand that I am a normal physical system and, as such, I am affected by the rest of reality. Hence I make a reasonable extrapolation, based on this and on my realisation that I am not special. I have no reason to believe that reality comes into being only when it interacts with me, and not also when anything interacts with anything else. That there is no fundamentally distinguished class of systems called ``the classical world'' or ``measuring apparata'' that have the privileged ability of actualising the variables of other systems.

Finally, Pienaar complains that the different ``views'' do not ``share facts''. Here, Pienaar puts  undue restrictions on what is a shared fact.

The analogy with conscious experience helps us here, too. Can two people ``share'' the same experience? It depends what we mean by that. If we mean to ask if two people can have the exact same set of sensory experiences at the same time and think the exact same thoughts, then clearly no. But this is not how we normally understand the phrase. We share experiences when we listen to the same performance of an orchestra, when we watch a movie together, when we analyse the same object together. And we can verify that we are sharing experiences by comparing our mental lives---not in some sort of absolute external sense, but by interacting with (talking and listening to)  each other. The two internal mental lives are still different after talking, but the two people can nevertheless reach an intersubjective agreement.

In the ontology of RQM, two systems \F and \W cannot share the same facts about a third system \S in the sense that whenever there is a quantum event for \F, there is also immediately a quantum event for \W. It is not even the case that a later interaction between \W and \F can make a previous quantum event between \S and \F a quantum event for \W. What they \textit{can} do, however, is verify that there is a consistency between their shared perspective, by interacting with (or measuring, as Pienaar puts it) each other. In this sense, \F and \W end up sharing facts: the behaviour of \F and \S that \W observes is coherent with the assumption that \F sees the same \S as \W does.

The fear that this destroys the coherence of the world or throws us into a solipsistic nightmare is similar to the fear that by setting the Earth in motion Copernicus challenged the stability of the houses built on Earth.  

Yet, quantum physics teaches us that \W could also interact with \F in a way that destroys \F account of their previous interaction with \S.  Is this surprising? Perhaps, but this is what quantum physics implies.

\subsubsection*{The loose frame loophole}

Pienaar also raises a concern regarding the ambiguous way in which some RQM literature talks about facts relative to different systems. This is a valid concern. It has already been echoed out by at least one other source \cite{stacey2021relationalist}. This ambiguity is a defect of the original literature.

Statements such as ``when two systems \F and \W interact with a system \S, the perspectives of \W and \F agree, and this can be checked in a physical interaction'', which can be found in the RQM literature, mean \emph{only} that \W can interact with \F's pointers and check that they were affected in  its interactions with \S in a way consistent with what \W  directly learns about \S. This is the content of \textbf{RQM:6$\star$} again. Obviously, \F can do the same with \W (\textbf{RQM:1}). 

This is normally left implicit whenever one talks about facts relative to different observers, assuming that it is clear enough to fill in the gaps. This is sometimes easier and sometimes harder. Indeed, Pienaar brings up \cite{dibiagio2021stable} as an example in which things are more complex than even the authors of the original paper realised. Let us look at this example in detail and make sure that the loophole is closed, as this can serve as an example for other situations.
  
The central point of \cite{dibiagio2021stable} is the definition of  a \textit{stable fact}. A fact relative to \F is said to be \textit{stable for} \W if classical probability calculus can be used to compute the probability of an event for \W using this fact relative to \F. More specifically, assume that, from \W's perspective, \F interacts with a variable $A$ of \S; According to \textbf{RQM:1$\star$} and \textbf{RQM:4$\star$}, this interaction may result in the value of $A$ to become a fact relative to \F, but no fact is established relative to \W.  However, the value of $A$ relative to \F is  considered \textit{stable for} \W if, in computing the probability for a variable $B$ to taking the value $b$ relative to \W in a subsequent interaction, we can write:
\begin{equation}\label{pr}
  P(b^\superW) = \sum_i P(b|a_i)P(a_i^\superF).
\end{equation} 

In the formula above, we have inserted superscripts to highlight that stability allows to mix perspectives. At an operational level, it allows to reason \textit{as if} there is epistemic uncertainty about the value of $A$ relative to \W, even though, ontologically, $A$ does not have an actual value relative to \W. The conditional probability
\begin{equation}\label{born}
  P(b|a_i) = |\braket{b}{a_i}|^2
\end{equation}
does not need superscripts: the transition probabilities are the main outputs of quantum theory and they define the probability of facts relative to a system, given other facts relative to the \textit{same} system. (In \cite{dibiagio2021stable}, we wrote $P(b^\superW|a_i^\superF)$, but, as Pienaar remarks, this notation is highly misleading; a better notation would have been $P(b^\superW|a_i^\superW)$. Better still to omit the superscript, since the transition probabilities given by the Born rule \eqref{born} are independent of the context \W.)

Pienaar expresses his doubts that a formula like \eqref{pr} can ever make sense in RQM, as it relates probabilities of facts relative to \textit{different} systems.  Indeed, while quantum theory allows the computation of the transition probabilities $P(b|a_i)$ as well as the probability $P(b^\superW)$ (given the state of $\S-\F$ relative to \W),  the quantities $P(a_i^\superF)$ do not have meaning in RQM, \textit{a priori}. But this is precisely the point of the definition. The $P(a_i^\superF)$ \textit{acquire} this meaning when the relation between $P(b^\superW)$ and $P(b|a_i)$ is given by \eqref{pr}. In other words, when the interaction between \S and \F (as described by \W) is such that \eqref{pr} holds for \textit{some} probability distribution $P(a_i^\superF)$, then $P(a_i^\superF)$ acquires the meaning of a probability distribution over the possible values of $A$---even though the value of $A$ is not a fact relative to \W. The value of $A$ might be a fact relative to \F, hence the superscript. 

 The reader is invited to consider the example in section~{1.2} of \cite{dibiagio2021stable}, also reported in the appendix \ref{app}.

As Pienaar remarks, the de-labelling is methodological.  Even when \eqref{pr} holds, there is no ontological identification of a fact relative to \F with a fact relative to \W. For all practical purposes, different systems in the same stability class act \textit{as if} they live in a macroreality of absolute facts and \textit{as if} they share facts.

\subsubsection*{No-go theorem 4}
\label{sec:nogo4}

This theorem is the second horn of the dilemma that No-go theorem 3 was supposed to offer. We grabbed the first horn, so we are not required to answer to this, but we will do anyway, because is another example of the mischaracterisation of the role of the quantum state.

The theorem is in a form of a trilemma:
\begin{quote}
      \textbf{Trilemma:} The propositions \textbf{P1} \& \textbf{P2} and the claim \textbf{RQM:3} cannot all be true.
\end{quote}
where
\begin{quote}
    \textbf{P1.} \W can measure $\mathcal{F}$-$\mathcal{S}$ in any basis at Event 2, independently of which basis $\mathcal{F}$ measured $\mathcal{S}$ at Event 1.
\end{quote}
and
\begin{quote}
    \textbf{P2:} Suppose $\mathcal{W}$ measures $\mathcal{F}$-$\mathcal{S}$ in the $\{ \ket{Fy_m} \ket{y_n} \}$ basis and obtains some outcome, updating the state relative to $\mathcal{W}$ to one of the states in $\{ \ket{Fy_n}\ket{y_n} \}$ just after Event 2. Then we can interpret this state as indicating that `$\mathcal{F}$ measured $\mathcal{S}$ in the $\{ \ket{y_n} \}$ basis and obtained one of the outcomes in the set $\{ y_n \}$ at Event 1'.
\end{quote}
\begin{equation}
  \ket{\Psi}_{\S\F} = \sum_i   \alpha_{i}   \ket{x_i}_{\S} \ket{Fx_{i}}_{\F},
\end{equation}
The solution is simple: \textbf{P1} and \textbf{RQM:3} are true in RQM, while \textbf{P2} is false. This is again caused by the wrong formulation of \textbf{RQM:5}. The fact that $\S-\F$ is in the state $\ket{y_n}_\S\ket{Fy_n}_\F$ does not imply that the value of $Y$ is a fact for $\F$.
Indeed, one way to prepare such a state is to start with \S and \F uncorrelated and just rotate each system separately into $\ket{y_n}_\S\ket{Fy_n}_\F$. In this case \S and \F never interacted and there could not be a fact about \S relative to \F. Or, in the operational language, \F did not measure \S.

\subsubsection*{No-go theorem 5}
\label{sec:nogo5}

This last no-go theorem again fails because of Pienaar's wrong formulation of \textbf{RQM:5}.
The theorem considers particular states of the $\S-\F$ system and tries to derive something about facts of \S relative to \F from these states. Again, this is not a possible logic in RQM.   The states in question (as Pienaar himself points out) are  states relative to \W.  What they contain is information about what \W can measure, namely how the $\S-\F$ system has affected \W or can affect \W is the future.  Trying to read out from these states the full facts relative to \F is is not something compatible with RQM.

\section{Qubits are not observers}
\label{sec:qubits}

Let us now come to Brukner's no-go theorem \cite{brukner2021qubits}. Like Pienaar's results, this is a correct mathematical observation that instead of providing a criticism of RQM, serves to sharpen the interpretation. His explicit aim is presented in the introduction:
\begin{quote}
  I will derive a no-go theorem that restricts the possibility of understanding the relational description in RQM as knowledge that one system can have about another in the conventional sense of that term.
\end{quote}
Part of what makes Brukner's result seem a challenge towards RQM, is Brukner's use of operational language (such as ``measurement,'' ``observer,'' and ``knowledge'') to formulate his no-go theorem even though, as he himself remarks, ``RQM makes not reference to [these] concepts''.

The other aspect that contributes to the confusion is his overplaying the role of the quantum state. Like in Pienaar's no-go theorems 1, 2, 4, and 5, Brukner tries to read relative facts between two systems by looking at the state assigned to these by a third system, while RQM does not allow this. 

The setup of the theorem is essentially that of Pienaar's no-go theorem 1. Two systems\footnote{We switch from \S and \F to $S$ and $O$ to be closer to Brukner's~\cite{brukner2021qubits}.} $S$ and $O$ are in some potentially entangled state $\ket{\psi}_{SO}$. Note that (i) here $O$ stands for ``observer'', (ii) there is no restriction on the nature of $S$ and $O$ (they can be qubits), and (iii)  Brukner does not specify what the state $\ket{\psi}_{SO}$ is relative to, we take it as relative to a third system $W$. Then the theorem states that the two following things cannot both be true.
\begin{quote}
  \textbf{1. (DefRS) Definite Relative State} For \textit{any} set of states $\{\ket{x_i}_S,\ket{X_i}_O\}$ such that
\begin{equation}\label{psi-so}
\ket{\psi}_{SO} = \sum_i c_i \ket{x_i}_S\ket{X_i}_O,
\end{equation}
[...] the states $\ket{X_i}_O$ are states of knowledge of the observer. When the observer is in state $\ket{X_i}_O$ she knows that the system is in the definite relative state $\ket{x_i}_S$.
\\\\
 \textbf{2. (DisRS) Distinct Relative State} The observer’s states of knowledge $\ket{X}_O$ and $\ket{X}_O$, which are correlated with distinct relative states $\ket{x}_S$ and $\ket{x'}_S$ of the system, are represented by orthogonal vectors in the observer’s Hilbert space, i.e. if $\ket{x}_S  \neq \ket{x'}_S$, then $\braket{X}{X'} = 0$.
\end{quote}

 Since RQM makes no appeal to a notion of knowledge, it's not clear why this should be a challenge to RQM. From RQM's perspective, Brukner's result ostensibly is a no-go theorem about the meanings that the word ``knowledge'' can assume, given what we know about quantum mechanics. 
 
  Indeed, we see two ways of reading this result, either
\begin{itemize} \setlength\itemsep{-.3em}
  \item \textbf{DefRS} is taken as a \textit{definition} of the word ``knowledge,'' and then \textbf{DisRS} is false, or
  \item \textbf{DisRS} is a constraint on what can be a ``state of knowledge'', and then  \textbf{DefRS} is false.
\end{itemize}

If we take \textbf{DefRS} to define knowledge, then a $O$ has knowledge about $S$ in the sense that at that $W$ can learn about the probabilities of future interaction with $S$ by interacting with $O$. This is the same well-defined sense in which a given set of pixels on my computer screen have knowledge about the time and a given set of ink molecules in a book have knowledge about lasers: by interacting with those molecules I expect to learn about future interactions with coherent light. In this sense, ``knowledge'' is nothing more---and nothing less---than correlations between two systems, as expected by a third. Then the failure of \textbf{DisRS} tells us nothing we didn't know already: when $S$ and $O$ are entangled, interacting with different variables of $S$ affects our information about different variables of $O$.

Consider now using \textbf{DisRS} as a constraint on what is a state of knowledge. Then the failure of \textbf{DefRS} implies that one has to have correlations on a preferred basis before talking about knowledge. This is closer to the other meaning of the word ``knowledge''  as applied to complex systems such as agents and conscious observers. This is perhaps ``the conventional sense'' that Brukner's has in mind in the introduction, although it's still a naturalistic use of the world, as it refers to the physical properties of such observer. Then a superposition of two states of knowledge is not a new state of knowledge, but a superposition of two states of knowledge.  

Both choices are valid. The only problem is confusing the different possible meanings of the word knowledge. And that is why the RQM literature warns against using terms that are normally reserved for macroscopic physics when talking about the fundamental elements of the theory.

In sum, we agree with Brukner that ``qubits are not observers,'' for the uncontroversial fact that qubits are not decision-making agents capable of processing information. That has never been a claim of RQM.
The controversial claim that RQM makes is \textbf{RQM1$\star$ (facts happen relative to any physical system)}. Brukner's no-go theorem has no impact on this, since a state such as \eqref{psi-so} that $W$ assigns to $S$ and $O$ is not enough for $W$ to infer what might or might not be a fact for $O$, as explained in detail in section \ref{sec:nogo1} when discussing Pienaar's no-go theorem 1.

\section{Conclusion}
\label{sec:conclusion}

Pienaar's \cite{pienaar2021quintet} presents arguments against two ideas: that (i)  RQM preserves certain classical relativistic intuitions about relations and (ii) it preserves the idea that consistency can be established between different observers' accounts. 
Both conclusions are correct:  (i) RQM does not \emph{preserve} certain classical relativistic intuitions about relations: it \emph{extends them} and makes them more radical (``facts are relative''). And, (ii) RQM does not preserve the idea that consistency can be established between different observers' accounts. It replaces it with the idea that systems communicate in the sense that they can measure (quantum mechanically!) each other's pointer variables. Since I myself am an observer, I find nothing strange in the idea that you could read pointer variables in me that gets correlated with external variables when these are realised with respect to me. 

Brukner's \cite{brukner2021qubits} argues that if we want to call the entanglement of two systems  ``knowledge'' that the two systems have about one another, then this ``knowledge'' differs in some radical way from common usage. 

These objections do not challenge the coherence of RQM.  They maybe show that RQM is more radical than what it might appear at a first sight. 

Does an ontology where views cannot be compared directly and physical systems can only check each other via quantum measurements imply solipsism?  No, it does not. Does it change in depth our way of thinking about reality?  Yes it does.   Quantum mechanics is radical.  One way or the other, we have to embrace it, not try to tame it.

\begin{acknowledgements}
We thank Jacques Pienaar and \v{C}aslav Brukner for cordial and fruitful conversations about their recent works and the interpretations of quantum mechanics in general.  Thanks also to Wolfgang Wieland for an early exchange on Brukner's paper. 

This work is partly funded by the ID\# 61466 grant from the John Templeton Foundation, as part of the ``The Quantum Information Structure of Spacetime (QISS)'' Project (\href{qiss.fr}{qiss.fr}). 
\end{acknowledgements}

\vfill

\appendix

\section{How decoherence stabilises facts\footnote{This section is a direct adaptation of section 1.2 of \cite{dibiagio2021stable}.}}
\label{app}

Consider two systems $\S$ and $\E$ ($\E$ for ``environment''), and a variable $A$ of the system $\S$ to which \E is couple. Let $a_i$ be its eigenvalues. A generic state of the compound system $\S-\E$, relative to a third system \W, can be written in the form 
\begin{equation}\label{psi}
\ket{\Psi} = \sum_i \alpha_i \ket{a_i}\otimes\ket{\psi_i}, 
\end{equation}
where $ |\psi_i\rangle$ are normalised states of $\E$.  Let us define
\begin{equation}\label{epsilon}
\epsilon = \max_{i\neq j} |\braket{\psi_i}{\psi_j}|^2.  
\end{equation}
Now, suppose that: (a) $\epsilon$ is vanishing or very small and (b) $\W$ does not interact with $\E$. Then the probability $P(b^\superW)$ of any possible fact relative to $\W$ resulting from an interaction between $\S$ and $\W$ can be computed from the density matrix obtained tracing over $\E$, that is,
\begin{equation}
\rho= \tr_{\E} \ketbra{\Psi} = \sum_i \ |c_i|^2 \ketbra{a_i}+O(\epsilon) .
\end{equation} 
Note that, by posing $P(a_i^{\superE}) = |c_i|^2$, we can then write
\begin{equation}\label{almost}
P(b^{\superW}) = \sum_i P(b^{\superW}|a_i^{\superE})P(a_i^{\superE}) + O(\epsilon).
\end{equation}
Thus, probabilities for facts $b$ relative to $\W$ calculated in terms of the possible values of $A$ satisfy \eqref{pr}, up to a small deviation of order $\epsilon$. Hence, by our definition, the value of the variable $A$ is stable for $\W$---to the extent to which  one ignores effects of order $\epsilon$. 
In the limit $\epsilon\rightarrow0$, the variable $A$ of the system $\S$ is exactly stable for $\W$.   

\bibliography{pienaar}

\end{document}